\documentclass[prl,twocolumn,showpacs,amsmath,amssymb]{revtex4}

\usepackage{graphicx}
\usepackage{dcolumn}
\usepackage{bm}

\begin{document}

\title{Microwave-Induced Dephasing in One-Dimensional Metal Wires}

\author{J. Wei, S. Pereverzev and M. E. Gershenson}
\affiliation{%
Department of Physics and Astronomy, Rutgers University,
Piscataway, NJ 08854
}%

\date{\today}

\begin{abstract}
We report on the effect of monochromatic microwave (MW) radiation
on the weak localization corrections to the conductivity of
quasi-one-dimensional (1D) silver wires. Due to the improved
electron cooling in the wires, the MW-induced dephasing was
observed without a concomitant overheating of electrons over wide
ranges of the MW power $P_{MW}$ and frequency $f$. The observed
dependences of the conductivity and MW-induced dephasing rate on
$P_{MW}$ and $f$ are in agreement with the theory by Altshuler,
Aronov, and Khmelnitsky \cite{Alt81}. Our results suggest that in
the low-temperature experiments with 1D wires, saturation of the
temperature dependence of the dephasing time can be caused by an
MW electromagnetic noise with a sub-pW power.

\end{abstract}


\maketitle

The processes of dephasing of electron wave function are central
to the electronic transport in mesoscopic systems \cite{Alt85}.
The dominant low-temperature dephasing mechanism in
low-dimensional conductors is the scattering of an electron by
equilibrium fluctuations of the electric field in the conductor,
i.e. the Nyquist (Johnson) noise \cite{Alt82}. The Nyquist
dephasing time $\tau_{\varphi}$ increases with decreasing
temperature as $T^{-2/3}$ in the quasi-one-dimensional (1D)
conductors with the cross-sectional dimensions much smaller than
the dephasing length $L_{\varphi}=\sqrt{D\tau_{\varphi}}$ ($D$ is
the electron diffusion constant) \cite{Alt82}. In the 1D metallic
wires, this mechanism typically governs the dephasing at $T < 1$K
\cite{Ech93, Pie03, Nat01}.

Recently, the interest in the fundamental limitations on
$\tau_{\varphi}$ was invigorated by the reports on the saturation
of $\tau_{\varphi}(T)$ dependences in one- and zero-dimensional
systems at ultra-low temperatures (see, e.g. \cite{Moh97, Hui99}).
The experiments \cite{Pie03} demonstrated that, at least in some
studied 1D wires, this saturation could be attributed to the
presence of localized spins in a small concentration undetectable
by analytical methods. However, the problem of $\tau_{\varphi}(T)$
saturation in the most ``clean'' samples remained open
\cite{Moh03}. One of the``extrinsic'' mechanisms that might lead
to the saturation of $\tau_{\varphi}(T)$ is the dephasing by the
external microwave (MW) electromagnetic noise \cite{Alt81}.
Detection of a very weak MW noise that is sufficient to destroy
the phase coherence at ultra-low temperatures is a challenge.
Indeed, the MW-induced dephasing may occur without an
easily-observable electron overheating if the electrons in a wire
can efficiently dissipate their energy in the environment
\cite{Kha98}. The possibility of
``noise-dephasing-without-overheating'' has not been ruled out in
the experiments \cite{Moh97, Moh03, Web99}.

\begin{figure}
\vspace{-0.2in}
\includegraphics[width=3.4in]{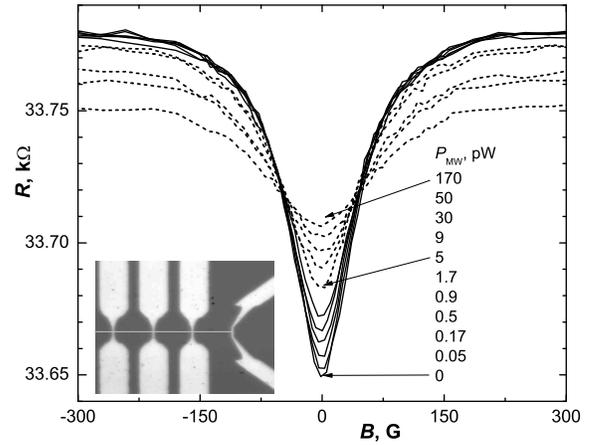}
\vspace{-0.35in}
 \caption{\label{fig1} WL magnetoresistance
measured at $T =0.2$K for different values of the microwave power
$P_{MW}$ of radiation with $f = 1$GHz. The solid lines correspond
to the power range $P_{MW}<5\cdot10^{-12}$W where electron
overheating is negligible; the dashed lines - to the power range
$P_{MW}>5\cdot10^{-12}$W where electron overheating by MW
radiation becomes significant. The microphotograph of a portion of
the sample is shown in the inset. } \vspace{-0.2in}
\end{figure}

In this Letter, we study the dephasing by monochromatic microwave
radiation in 1D wires. By optimizing the sample design, we
minimized electron overheating and observed for the first time the
microwave-induced dephasing in 1D wires without a concomitant
overheating of electrons. The dependences of the MW-induced
dephasing rate on the MW power and frequency are in agreement with
the theoretical predictions \cite{Alt81}. Our results suggest that
the $\tau_{\varphi}(T)$ dependence in a 1D wire may be
significantly affected by a sub-pW power of an external microwave
noise absorbed in the sample.

The challenging aspect of the experiments on MW-induced dephasing
is the separation of this effect from a trivial MW-induced
electron overheating, which also leads to the dephasing
enhancement \cite{Wan87,Liu89,Vit88}. An efficient cooling of the
electrons is crucial for this separation. The amplitude of the MW
electric field $E_{MW}$ that leads to a strong MW-induced
dephasing within the optimal frequency range $\omega
\tau_{\varphi} \sim 1$ can be estimated from the condition
$eE_{MW}L_{\varphi}\sim \frac{\hbar}{\tau_{\varphi}}$
\cite{Alt81}. The corresponding MW power is \emph{proportional} to
the wire length $L$:
\begin{eqnarray}\label{P-dephasing}
    P_{\varphi}\equiv(E_{MW}L)^2/R=\frac{\hbar^{2}L}{D\tau_{\varphi}^{3}R_{1}e^{2}}.
\end{eqnarray}
Here $R_1=R/L$ is the resistance of the wire per unit length. For
observation of the MW-dephasing-without-overheating, dissipation
of this MW power in the wire should not affect significantly the
electron temperature $T_e$. The electrons in 1D wires transfer
their energy to the environment via (a) the electron-phonon
scattering (which becomes very weak at $T<1$K \cite{Ger01}) and
(b) the outdiffusion of hot electrons into cooler current leads.
For the latter mechanism, the dissipated power \cite{Pro93}
\begin{eqnarray}\label{P-es}
    P_{diff}=(\frac{2\pi k_B}{e})^{2} \frac{1}{2R_{1}L}
    (T_{e}^{2}-T^{2})
\end{eqnarray}
is \emph{inversely proportional} to $L$. Thus, in
\emph{sufficiently short} wires, $P_{\varphi}$ can be dissipated
without overheating. However, short wires with a small total
resistance are more susceptible to the overheating by external
electromagnetic noise. Also, a large amplitude of universal
conductance fluctuations in short wires reduces the accuracy of
extraction of $\tau_{\varphi}$ from the weak localization (WL)
magnetoresistance. We resolved this dilemma by attaching the
cooling fins \cite{Kau98} to a long ($L$=1200 $\mu$m) wire (see
the inset in Fig.~\ref{fig1}). These fins provide the heat sinks
for the hot electrons in the wire and improve significantly the
electron cooling \cite{heating}. At the same time, the cooling
fins do not affect the 1D WL correction to the conductivity
provided the spacing between them ($L^*$=30 $\mu$m in our samples)
is much greater than $L_{\varphi}$ ($\sim4.5\mu$m at $T=0.05$K for
the studied wires) \cite{Ger95}.

\begin{figure}
\vspace{-0.2in}
\includegraphics[width=3.4in]{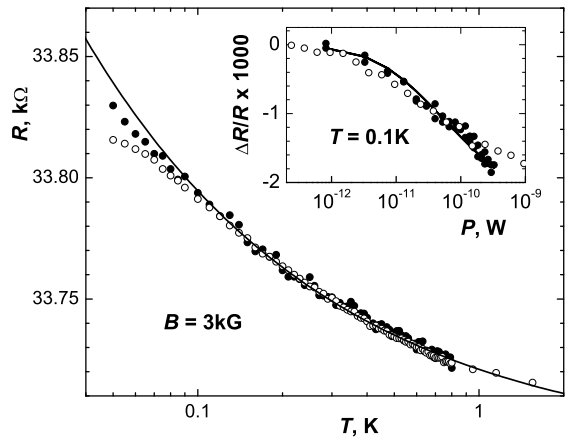}
\vspace{-0.35in} \caption{\label{fig2} Dependences $R(T,B=3$kG$)$
measured at $I_{ac}=$ 3nA({\Large $\bullet$}) and 10nA({\Large
$\circ$}). The solid line corresponds to the EEI correction
$\Delta R_{EEI}(T)$ (Eq.\ref{EEI}). The inset shows the procedure
of calibrating the MW power dissipated in the wire (see the text):
the dependences $R(P_{dc})$ ({\Large $\bullet$}) and $R(P_{MW})$
({\Large $\circ$}) were measured at $T=0.1$K and $B=3$kG. The MW
power was found by matching $R(P_{MW})$ with $R(P_{dc})$. The
solid line shows $\Delta R_{EEI}[T_e(P)]$ (Eq.\ref{EEI}) where the
electron temperature $T_e(P)$ was calculated from Eq.\ref{P-es}.
We found experimentally that an additional prefactor $\sim 0.2$
should be added to the right-hand side of Eq.\ref{P-es} to account
for a less-than-perfect efficiency of cooling fins \cite{heating,
calibration}. } \vspace{-0.2in}
\end{figure}

The 1D silver wires were fabricated by e-beam lithography and
thermal evaporation of Ag (purity 99.9999\%). Similar results have
been obtained for several samples; below the data are presented
for a wire with the width $W = 69$ nm, thickness $d = 20$ nm and
the diffusion constant $D = 110$ cm$^{2}$/s. The resistance was
measured by the \emph{ac} resistance bridge at the frequency 13
Hz over the temperature range $0.05-1$K. Special care was taken
to reduce the MW noise level in the dilution refrigerator by
installing low-pass filters in all lines at the top of the
cryostat and at the cold finger near the sample. The central and
outer conductors of the broad-band ($f = 0-26$ GHz) coaxial cable
were coupled to the sample via 5-nF DC block capacitors. For
suppressing the external MW noise and room-temperature thermal
radiation, the cable was interrupted by two 20-dB attenuators at
$T=4$K and 1K.

The evolution of the WL magnetoresistance with $P_{MW}$, measured
at a fixed bath temperature $T=0.2$K, is shown in Fig.~\ref{fig1}.
The positive magnetoresistance observed in weak magnetic fields
for the studied wires is due to the field-induced suppression of
the WL corrections to the resistivity in the presence of strong
spin-orbit scattering (the so-called weak antilocalization, see,
e.g., \cite{Alt87}). Over a wide range of $P_{MW}$ ($\sim20 dB$),
the only observable change in the MR is a decrease of the
amplitude of the $B=0$ ``dip'' associated with an increase of the
dephasing rate. Within this $P_{MW}$ range, the dependences
$R(B)$ outside the dip are not affected by radiation, which
indicates that the electrons remain in equilibrium with the bath.
The electron overheating at $P_{MW}>10^{-11}$W leads to the
decrease of $R$ in strong magnetic fields ($L_{H}\ll
L_{\varphi}(T)$, where $L_H= \sqrt{\hbar/2eH}$ is the magnetic
length), which is associated with the temperature dependence of
the electron-electron interaction correction to the conductivity
\cite{Alt85}:
\begin{eqnarray}\label{EEI}
    \frac{\Delta R_{EEI}(T)}{R}=\frac{e^{2}L_{T}R_{1}}{\pi \hbar},
 \end{eqnarray}
where $L_{T}=\sqrt{\hbar D/k_{B}T}$.

We have used the measurements of $\Delta R_{EEI}(T,P)$
(Fig.~\ref{fig2}) for an estimate of the electron temperature
$T_e$ \cite{Te}. The dependence $\Delta R_{EEI}(T)$ was also used
for calibration of the microwave power dissipated in the wire. In
these measurements the electrons were overheated above a fixed
bath $T$ either by the \textit{dc} current $I_{dc}$ or by the MW
radiation. The decrease of the resistance, $\Delta R$, was
recorded (a) as a function of the \textit{dc} power
$P_{dc}=I_{dc}^{2}R$ at $P_{MW} = 0$ and (b) as a function of
$P_{MW}$ at $P_{dc}$ = 0 (see the inset in Fig.~\ref{fig2}).
Assuming that the heating by the \textit{dc} current is the same
as that by MW radiation in the limit $f \ll (2\pi \tau )^{-1}$
($\tau \sim 10^{-15}$ s is the momentum relaxation time in our
wires), one can estimate the MW power dissipated in the sample
\cite{calibration}.

\begin{figure}
\vspace{-0.2in}
\includegraphics[width=3in]{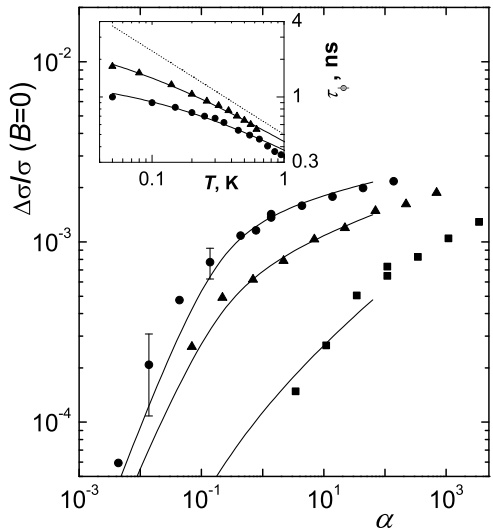}
\vspace{-0.2in} \caption{\label{fig3} Dependences of the
conductivity $\triangle\sigma/\sigma (B=0)$ on the normalized MW
power $\alpha$, measured at 0.5 K at different frequencies [1GHz
({\Large $\bullet$}), 0.5GHz ($\blacktriangle$), 0.2GHz
($\blacksquare$)]. The theoretical dependences
(Eq.~\ref{correction}) calculated for these frequencies are shown
by the solid curves. The inset shows the dependences
$\tau_{\varphi}(T)$ measured with no intentionally applied
monochromatic MW radiation for the wire coupled to the coaxial
cable with one ``cold'' 20dB attenuator ({\Large $\bullet$}) and
two ``cold'' 20dB attenuators connected in series
($\blacktriangle$). The dashed line shows the $T$-dependence of
the Nyquist dephasing time (Eq.~\ref{tau-phi}), the solid curves
- fitting by Eq.~\ref{comb-rate} with $\tau_0=1.5$ns (3.8ns) for
20dB (40dB) attenuation.} \vspace{-0.2in}
\end{figure}

The dephasing time $\tau_{\varphi}$ was extracted from the
magnetoresistance (MR) using the theoretical expression for the
1D WL MR modified for the case of strong spin-orbit scattering
($\tau_{SO} \ll \tau_{\varphi}(T)$) \cite{Alt82, Ale99}:
\begin{eqnarray}\label{WL}
    \frac{\Delta R_{WL}}{R}=\frac{e^{2}L_{\varphi}R_{1}}{\pi \hbar}
    [\frac{3}{2}\frac{Ai(\tau_{\varphi}^{*}/\tau_H)}{Ai(\tau_{\varphi}^{*}/\tau_H)'}
    -\frac{1}{2}\frac{Ai(\tau_{\varphi}/\tau_H)}{Ai(\tau_{\varphi}/\tau_H)'}].
\end{eqnarray}
Here $\tau_{H}=\frac{12L_{H}^4}{DW^2}$,
$(\tau_{\varphi}^{*})^{-1}=\tau_{\varphi}^{-1}+\frac{4}{3}\tau_{SO}^{-1}$,
$Ai(x)$ is the Airy function. The temperature dependences of
$\tau_{\varphi}$ measured with no intentionally applied
monochromatic MW radiation are shown on the inset in
Fig.~\ref{fig3}. These dependences can be fitted with the
expression
\begin{eqnarray}\label{comb-rate}
    \tau_{\varphi}^{-1}=\tau_{\varphi0}^{-1}+\tau_{0}^{-1} ,
 \end{eqnarray}
where $\tau_{\varphi0}$ is the Nyquist dephasing time \cite{Alt82,
Ale99}
\begin{eqnarray}\label{tau-phi}
    \tau_{\varphi
    0}=(\frac{\hbar^2}{e^{2}k_{B}TR_{1}\sqrt{D}})^{\frac{2}{3}}.
 \end{eqnarray}
We have observed that the $T$-independent ``cut-off'' term
$\tau_0$ increased from 1.5 ns to 3.8 ns with the attenuation in
the MW line being increased from 20dB to 40dB. Presumably, this
$\tau_0$ increase reflects suppression of the MW noise delivered
to the sample by the MW line (see below).

On Fig.~\ref{fig3}, we compare the observed $P_{MW}$-induced
increase of $\sigma (B=0)$ with the theoretical prediction
\cite{Alt81}:
\begin{eqnarray}\label{correction}
    \Delta\sigma=\frac{2e^{2}\sqrt{D}}{\hbar \sqrt{\pi \omega}}\!
    \int_{\omega \tau}^{\infty}\!\frac{dx}{\sqrt{x}}\exp[-\alpha f(x)-\frac{2x}{\omega
    \tau_{\varphi}}]
     \,I_{0}[\alpha f(x)]
 \end{eqnarray}
Here $\alpha \equiv
\frac{2e^{2}D(E_{MW})^{2}}{\hbar^{2}\omega^{3}}$, $\omega=2\pi f$
is the angular frequency of MW radiation, $I_{0}$ is the Bessel
function of an imaginary argument. In calculating the theoretical
dependences $\Delta\sigma(\alpha,\omega,\tau_{\varphi})$ in
Fig.~\ref{fig3}, we took into account that the electron
overheating at large $P_{MW}$ (e.g., $P_{MW}>10^{-11}$ W at $T$ =
0.2 K) leads to the decrease of $\tau_{\varphi}$:
$\tau_{\varphi}(P_{MW})$ was calculated from the measured
dependences $\tau_{\varphi}(T,P_{MW} = 0)$ (the inset in
Fig.~\ref{fig3}) and $T(P_{MW})$ (Fig.~\ref{fig2}). Our
experimental data are in quantitative agreement with the theory
\cite{Alt81} over a broad range of $P_{MW}$. Note that after the
electric field in the wire has been determined experimentally, no
fitting parameters are involved in the comparison between the
data and the theory.

\begin{figure}
\vspace{-0.1in}
\includegraphics[width=3.4in] {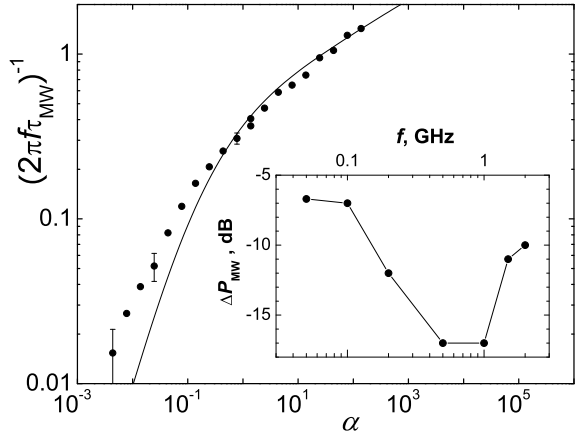}
\vspace{-0.3in} \caption{\label{fig4} Dependence of the MW-induced
dephasing rate $\tau_{MW}^{-1}$ on $\alpha$, measured at $T =0.5$
K at $f=1GHz$. Equation \ref{tau-MW1} is shown by the solid curve.
The insert shows the range of $P_{MW}$ where the
``dephasing-without-heating'' was observed at $T = 0.2$ K.}
\vspace{-0.2in}
\end{figure}

A more intuitive (though less rigorous \cite{reason}) way to
interpret the MW-induced change in the WL contribution is to
associate it with the MW-induced increase of the dephasing rate
(see, e.g., \cite{Web99,Wan87}):
\begin{eqnarray}\label{tau-MW}
   \tau_{MW}^{-1}(P_{MW})=\tau_{\varphi}^{-1}(T_{e},P_{MW})-\tau_{\varphi}^{-1}(T_{e}).
 \end{eqnarray}
Here $\tau_{\varphi}^{-1}(T_{e})$ and
$\tau_{\varphi}^{-1}(T_{e},P_{MW})$ are the dephasing rates at
$P_{MW}=0$ and $P_{MW}\neq 0$, respectively. When the electron
overheating becomes significant at large $P_{MW}$,
$\tau_{\varphi}(T_e)$ rather than $\tau_{\varphi}(T)$ should be
used in Eq.~8. Figure~\ref{fig4} shows the dependence of the
MW-induced dephasing rate on the normalized MW power $\alpha$ at
$f=1GHz$. It is worth mentioning that for the studied wires with
optimized electron cooling, the difference between the values of
$\tau_{MW}^{-1}(P_{MW})$ calculated with $\tau_{\varphi}(T_e)$ and
$\tau_{\varphi}(T)$ remains small even at high MW power levels:
e.g., this difference does not exceed $30\%$ at $\alpha=50$
($P_{MW} = 10^{-8}$ W at $f=1$GHz). The dependence $(2\pi
f\tau_{MW})^{-1}(\alpha)$ shown in Fig.~\ref{fig4} is in good
agreement with the estimate for $\tau_{MW}^{-1}(P_{MW})$ obtained
for optimal frequency $f\sim\tau^{-1}_{\varphi}$ at
$\tau_{MW}(P_{MW})\ll\tau_{\varphi}(T_{e})$ \cite{Alt81}:
\begin{eqnarray}\label{tau-MW1}
   \tau_{MW}(P_{MW})=\omega^{-1}\left\{
    \begin{array}{ll}
        (45/2\alpha)^{1/5}, \alpha\gg1 \\
        \alpha^{-1}, \alpha\ll1.
    \end{array}\right.
 \end{eqnarray}
In particular, at $\alpha\gg1$, the observed dependences approach
the asymptotic behavior $(2\pi f\tau_{MW})^{-1}\propto
\alpha^{1/5}$.

The inset in Fig.~\ref{fig4} shows how the range of $P_{MW}$,
where the MW-induced dephasing is observed without electron
overheating, depends on the MW frequency at $T=0.2K$. $\Delta
P_{MW}$ represents the ratio of $P_{MW}$ that causes a measurable
increase of dephasing rate (5 $\Omega$ increase of $R$ at $B=0$,
see Fig.~\ref{fig1}) to $P_{MW}$ that caused a noticeable increase
of $T_{e}$ (5 $\Omega$ decrease of $R$ at $B=3$ kG). Note that
the ratio is independent of the (frequency-dependent) coupling of
the wire to MW radiation. The
``MW-induced-dephasing-without-overheating'' was observed over
$\sim 1.7$ decades of $P_{MW}$ within the range $f = 0.5-1$ GHz.
``Shrinking'' of the $\Delta P_{MW}$ range for both higher and
lower $f$ is consistent with the prediction \cite{Alt81} that the
MW-induced dephasing is most efficient at $f \sim
\tau^{-1}_{\varphi 0}(T)$. Note that the characteristic frequency
dependence of the observed effect clearly distinguishes it from
the \emph{dc}-bias-driven ``dephasing-without-overheating'' in
the films with a high density of two-level systems \cite{zvi}.

An increase of $\tau_{0}$ with increasing attenuation in the MW
line (see the inset in Fig.~\ref{fig3}) suggests that the
saturation of $\tau_{\varphi}(T)$ observed in our experiment
below 0.2 K may be attributed to the dephasing by the external MW
noise. Quantitatively, the upper bound on the microwave noise
power dissipated in the wire is of the order of $10^{-13}$ W when
the wire is connected to the coaxial cable with two 20-dB
attenuators (this power is equivalent to $P_{ac}(I_{ac}=3nA)$, see
Fig.~\ref{fig2}). Assuming that the MW noise spectrum is peaked
within the frequency range most efficient for dephasing
($f\sim\tau_\varphi^{-1}\sim1$GHz), one can estimate the
noise-induced ``cut-off'' of the dephasing time $\sim 10$ ns.
This cut-off is close to the value of $T$-independent term
$\tau_{0}$ in Eq.\ref{comb-rate}. Thus, we conclude that the
saturation of dephasing time observed in our experiment at $T
\leq$ 0.1 K may be caused by an insufficient screening of the
sample from the external microwave noise, including the Nyquist
noise from all elements of the measuring set-up.

In summary, we observed for the first time the microwave-induced
dephasing in 1D metal wires without a concomitant overheating of
the electrons. The key requirement for observation of this effect
is the optimization of electron cooling in 1D wires. The observed
dependences of the weak-localization correction to the
conductivity on the microwave power and frequency are in a
quantitative agreement with the theory \cite{Alt81}. Our
experiments demonstrate that an ultra-low-noise environment is
essential for the experiments on fundamental limits of dephasing
time at low temperatures.

We thank B. Altshuler and V. Fal'ko for useful discussions. The
work was supported in part by the Rutgers Academic Excellence Fund
and the NSF grant DMR-0508129.


\begin{thebibliography}{99}
\bibitem{heating} The power that can be dissipated in our samples without overheating is smaller
 by a factor 2-5 than that predicted by Eq.~\ref{P-es} for a 1D wire flanked by infinitely large contacts.
\bibitem{Te} Comparison of Fig.~1 and Fig.~2 shows
that at $T=0.2$ K and $P_{MW} = 1.7\times 10^{-10}$ W, $T_e$ in
the wire is $\sim 0.6$ K. Even at the maximum $P_{MW} = 10^{-8}$ W
the overheating did not exceed 1.2 K at the bath $T = 0.5$ K.
\bibitem{calibration} At large $P_{MW}$ values, the wire
temperature may be affected by the MW power dissipation in the RC
filters installed close to the sample. Thus, our calibration
procedure provides the upper bound for $P_{MW}$ and $E_{MW}$ in
the wire.
\bibitem{reason} As it can be seen from Eq.~\ref{correction}, the effect of MW
radiation on the WL correction cannot be fully accounted by
modification of the dephasing rate.
\end{thebibliography}
\end{document}